\documentclass[aps,pre,twocolumn,groupedaddress]{revtex4-1}

\usepackage{graphicx}
\usepackage{amsmath}
\usepackage{amsfonts}
\begin{document}


\newcommand{\comment}[1]{} 
\newcommand{\be}{\begin{equation}} 
\newcommand{\ee}{\end{equation}}
\newcommand{\mat}[1]{{#1}}
\newcommand{\ba}{\begin{align}} 
\newcommand{\ena}{\end{align}}
\newcommand{\exval}[1]{\langle {#1} \rangle}
\newcommand{\bi}{\begin{itemize}} 
\newcommand{\ei}{\end{itemize}} 
\newcommand{\KM}{\mat{G}} 
\newcommand{\km}{g} 
\newcommand{\ra}{y} 
\newcommand{\Ra}{\mat{Y}} 
\newcommand{\avgR}{\bar y} 
\newcommand{\rae}{y} 
\newcommand{\sk}{s} 
\newcommand{\CM}{\mat{C}} 
\newcommand{\cm}{c} 
\newcommand{\sw}{g} 
\newcommand{\rv}{V_{\theta}} 
\newcommand{\gest}{\hat{g}}
\newcommand{\FT}[1]{\mathcal{F}[#1]}
\newcommand{\eff}[1]{\tilde{#1}} 
\newcommand{\LT}[1]{\mathcal{L} [#1]} 


\title{Recurrent Interactions in Spiking Networks with Arbitrary Topology}


\author{Volker Pernice}
\email[]{pernice@bcf.uni-freiburg.de}
\author{Benjamin Staude}
\author{Stefano Cardanobile}
\author{Stefan Rotter}
\affiliation{Bernstein Center Freiburg \& Faculty of Biology University of Freiburg, Germany}


\date{\today}

\begin{abstract}
The population activity of random networks of excitatory and inhibitory leaky integrate-and-fire (LIF) neurons has been studied extensively. In particular, a state of asynchronous activity with low firing rates and low pairwise correlations emerges in sparsely connected networks. We apply linear response theory to evaluate the influence of detailed network structure on neuron dynamics. It turns out that pairwise correlations induced by direct and indirect network connections can be related to the matrix of direct linear interactions.
Furthermore, we study the influence of characteristics of the neuron model. Interpreting the reset as self-inhibition we examine its influence, via the spectrum of single neuron activity, on network autocorrelation functions and the overall correlation level. The neuron model also affects the form of interaction kernels and consequently the time-dependent correlation functions.
We finally find that a linear instability of networks with Erd\H{o}s-R\'{e}nyi topology coincides with a global transition to a highly correlated network state. 
Our work shows that recurrent interactions have a profound impact on spike train statistics and provides tools to study effects of specific network topologies.
\end{abstract}


\maketitle

\section{Introduction}
 The dynamics of networks of interacting units, e.g.\ correlation and synchronization, depends on the properties of the nodes, as well as on the organization of their interactions. 
Neurons in cortical networks typically spike asynchronously  despite a large number of synapses coupling the neurons together. This behavior has been explained by random balanced networks of excitatory and inhibitory neurons \cite{vanVreeswijk1996, Brunel2000}. 
Later studies examined in greater detail how connections between neurons mediate correlations \cite{Kriener2008,*Shea-Brown2008,*Moreno-Bote2008}, but inhibitory network feedback decorrelates activity \cite{Hertz2010,*Renart2010,*Helias2011}. Residual correlations are shaped by the detailed structure of the network and potentially influence neuronal information processing.
Because the spiking operation typically incorporates a thresholding mechanism, single neuron models are intrinsically nonlinear. A further complication results from the fact that even in sparse networks recurrent interactions or feedback loops have a large influence on dynamics, leading to chaotic dynamics \cite{vanVreeswijk1996,Jahnke2009}. This implies that analytic results on a network level are difficult to obtain.
Here we use the impulse response to construct an efficient linearization in the asynchronous-irregular state of a network of LIF neurons.
Linear models represent a powerful tool to study structured networks in detail \cite{Cessac2004,*Galan2008,*Ostojic2009,*Lindner2005,Pernice2011}. 
As it turns out, effects of the network can be well described in such a linear framework. This provides a convenient way to include the full recurrent connectivity of the network and examine its interaction with single-neuron dynamics and demonstrates that the implications of recurrent feedback can be understood as a linear, albeit not trivial effect.

\section{Network model}
The LIF neuron is a standard model in studies of neural network dynamics. The dynamics of the membrane potential $V_k$ of neuron $k$ is governed by the equation
\be
\tau_m\dot{V_{k}}=-V_{k}-\tau_m\rv\sk_k+\tau_m\sum_jJ_{kj}\sk_j,
\ee
where $\tau_m$ denotes the membrane time constant and $J_{kj}$ the synaptic weight between presynaptic neuron $j$ and postsynaptic neuron $k$.  A spike is emitted when the membrane potential exceeds a threshold $V_T$. Spikes are modeled as $\delta$-shaped events, such that the spike-train of neuron $j$ is $s_j(t)=\sum_l\delta(t-t_l^{j})$. The membrane potential is decreased by an amount $\rv$ after each spike.
We simulate random networks comprising $N$ neurons of which $\epsilon = 80\,\%$ are  excitatory and $1-\epsilon = 20\,\%$ are inhibitory. Connections are established independently with probability $p=0.1$, such that a diluted network is obtained. Synaptic weights are  $J_E$ for excitatory and $J_I=-\gamma J_E$ for inhibitory neurons. A value of $\gamma>4$ then implies a relative dominance of inhibition in the network.
To study the system in a regime where neurons are mainly driven by membrane potential fluctuations, we assume that all neurons in the network receive Poisson-like input from external neurons with synaptic weights $J_E$ and rate $y_{ext}=V_T/(J_E\tau_m)$, such that the mean membrane potential is just high enough to reach threshold. 
In simulations we use $\tau_m=20\,\mathrm{ms}$, a refractory time  $\tau_r=2\, \mathrm{ms}$ after each spike, a threshold $V_T=20\,\mathrm{mV}$ and a reset by $\rv=20\,\mathrm{mV}$ if not mentioned otherwise. Network simulations were conducted with the NEST-simulator \cite{Gewaltig2007}.

\begin{figure}
\includegraphics[width=\columnwidth]{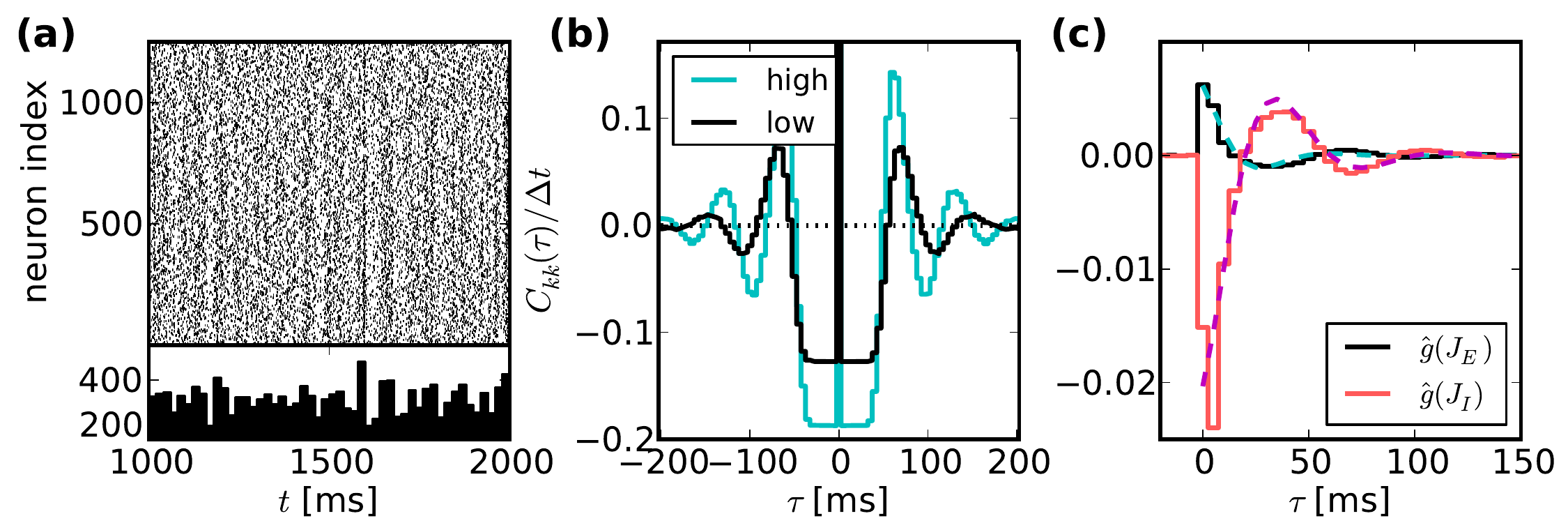}
\caption{ (color online) (a) Spike trains in a random LIF network ($N=1250,\gamma=5, J_E=0.1\,\mathrm{mV}$, simulation time $5\cdot 10^4\, \mathrm{s}$). Covariances and fluctuations of population spike counts (bottom) are small.
(b) Autocovariance of neurons with low and high rate. Normalization over bin size $\Delta t$ was chosen such that the integral of the peak corresponds to firing rate.
(c) Histogram of the response to external excitatory and inhibitory input spikes, averaged across neurons, as well as fits to a function $\cos(\nu \tau)e^{-\tau/\alpha}$ (dashed lines).
}
\label{netwState}
\end{figure}

The dynamic state of random networks of LIF neurons depends, besides the strength of the external input, on the ratio of excitation and inhibition. We consider neurons in balanced or inhibition dominated networks which spike asynchronously and irregularly at stationary firing rates, see Fig.\ \ref{netwState}(a). As a measure of correlations between spike trains we use the covariance density function
\be
\label{covEls}
\CM_{kj}(\tau)=\exval{\sk_k(t)\sk_j(t+\tau)}-\exval{\sk_k(t)}\exval{\sk_j(t)}.
\ee
 The autocovariances $C_{kk}$ exhibit troughs around $\tau=0$ because the reset after each spike leads to an effective dead time, panel (b).
The effect of a single input spike of weight $J$ on the firing rate of neuron $k$, the impulse response $\gest_k(J)$, panel (c), can be measured as the normalized covariance of some external Poisson input $s_e(t)$ coupled to $k$ with weight $J$,
\be
\label{impResp}
\gest_k(J,\tau)=\frac{\exval{s_e(t+\tau),s_k(t)}-\exval{s_e(t)}\exval{s_k(t)}}{\exval{s_e(t)}}.
\ee

\section{Linearly interacting point processes}

 A simple description of spike train dynamics is provided by the model of interacting point processes introduced by Hawkes \cite{Hawkes1971}.
Spike trains $\sk_k(t)$ are interpreted as realizations of Poisson processes with dynamic rates $\ra_k(t)=\exval{\sk_k(t)}$. The spike train vector $\sk(t)$ and the rate vector $\ra(t)$ are used to describe a population of neurons. External input is assumed to evoke a baseline firing rate  $\ra_0$. Presynaptic spikes at time $t-\tau$ cause a transient response. All interaction kernels are collected in the matrix $\KM(\tau)$, with $\KM(\tau)=0$ for $\tau<0$. 
The dynamics of the system is then defined by the equation
\be 
\label{Hawkesbasic} 
\ra(t)=\ra_0+\int_{-\infty}^{\infty}\KM(t-\tau)\sk(\tau)\,d\tau =\ra_0+\left(\KM\star\sk\right)(t).
\ee
In a stationary system, the time averaged rates are
\be 
\label{HawkesratesFirst} 
\ra=\left(1-\int_{-\infty}^{\infty}\KM(\tau)\,d\tau\right)^{-1} \ra_0,
\ee
where  $y=\langle s(t) \rangle$ was used. The matrix of crosscovariances $\CM(\tau)$ is determined by the self-consistency equation
\be
\label{selfConsEq}
\CM_0(\tau)=\KM(\tau)\Ra+(\KM\star\CM_0)(\tau),
\ee
where $\CM_0(\tau)=\CM(\tau)-\Ra\delta(\tau)$ denotes the continuous part of the covariance matrix. The rate matrix $\Ra$  is diagonal,  with elements $[\Ra]_{kj}=\ra_{k}\delta_{kj}$, coinciding with the integrated autocovariances of Poisson processes. This represents the fact that isolated nodes with no interactions are modeled as Poisson processes.
In the Fourier domain, the solution of \eqref{selfConsEq} is
\be
\label{fullCorrFuncs}
\CM(\omega)=[\mat{1}-\KM(\omega)]^{-1}\Ra[\mat{1}-\KM^*(\omega)]^{-1},
\ee
where $\KM^*$ denotes the transpose and complex conjugate of $\KM$, see  \cite{Hawkes1971} for details.
The reset after each spike affects the autocovariance functions of LIF neurons.
It is therefore not feasible to assume that isolated nodes spike as Poisson processes.
Nonetheless, their activity can be described in the Hawkes framework, if reset effects are taken into account. This can be done by interpreting the reset as self-inhibition.

In a network with synaptic coupling matrix $\KM_{syn}$, self-inhibition is described by diagonal elements $-\theta_{k}$, so that the full coupling matrix is $\KM=\KM_{syn}-\Theta$, with $[\Theta]_{kj}=\delta_{kj} \theta_k$.  In this case, Eq.\ \eqref{fullCorrFuncs} can be rewritten as
\be
\label{effCorrs}
\CM(\omega)=[\mat{1}-\eff{\KM}(\omega)]^{-1}\eff{\Ra}(\omega)[\mat{1}-\eff{\KM}^*(\omega)]^{-1}
\ee
with  $\eff{\KM}(\omega)=(1+\Theta(\omega))^{-1}\KM_{syn}(\omega)$ and  $\eff{\Ra}(\omega)=(1+\Theta(\omega))^{-1}\Ra(1+\Theta^*(\omega))^{-1}$.
One can interpret $\eff{\KM}$ as effective interaction matrix between neurons with reset. The elements of $\eff{\Ra}$ coincide with the autocovariances of univariate Hawkes processes with self-inhibition, which are more regular than pure Poisson processes. 
Covariances decrease, because the couplings  $[\KM_{syn}]_{kj}$ are weakened to $[\KM_{syn}]_{kj}/(1+\theta_k)$.

\section{Mapping to LIF networks} 
The prediction from the linear theory in \eqref{effCorrs} has been rewritten so that all quantities are experimentally accessible and can therefore be tested in simulations.
In a first step, we show how the linear response of neurons in a network can be identified with the elements of the effective coupling matrix $\eff{\KM}$. The coupling matrix determines the extent and distribution of covariances in the network.
In the second step, the effective autocovariance can be determined from measured autocovariances for a quantitative prediction of crosscovariance functions.

We shortly summarize some definitions at this point. The effective coupling matrix $\eff{\KM}$ results from reset effects in the actual coupling matrix $\KM$. External input with coupling $g_{ext}(\omega)$ to linear networks or synaptic weights $J_E,J_I$ to networks of LIF neurons can be used to infer an estimate $\gest(\omega)$ of the impulse response of network neurons. Integrated estimated impulse responses will be denoted by $\gest_E,\gest_I$ for excitatory and inhibitory neurons, respectively.

Analogously as in \eqref{impResp}, the impulse response $\gest_k(g_{ext})$  of neuron $k$ is measured as the covariance between the spikes of an external input node coupled to $k$ with kernel $g_{ext}$.
This covariance can be calculated using Equation \eqref{effCorrs}, where the coupling matrix has to be extended to include the external input. 
Specifically, this is done by calculating covariances in a network with coupling matrix $\KM'$ with entries $\KM'_{k1}=g_{ext}$, $\KM'_{k+1,j+1}=\eff{\KM}_{kj}$ for $k,j=1\dots N$ and 0 otherwise, such that the new node $
1$ corresponds to the external input.
The result is 
\be
\gest_k(g_{ext})=\frac{C'_{k1}}{y_{11}}=\frac{g_{ext}}{(1+\theta_k)}[(1-\eff{G})^{-1}]_{kk}.
\ee
Here, $y_{11}$ is the rate of the external input.
The factor $[(1-\eff{G})^{-1}]_{kk}$ relates to the fact that, as the neuron is
embedded in a network, the total response that is measured includes the direct response as well as the indirect response via its postsynaptic partners, which in turn project back to the original neuron.
For large diluted networks with $\exval{\eff{G}_{kj}}\ll 1$ this factor is negligible.
Therefore, $\gest_k(g_{ext})\approx g_{ext}/(1+\theta_k)$. If $g_{ext}$ is chosen identical to the internal couplings, measurement of the response then returns the elements of the effective coupling matrix. 
In simulated networks of LIF neurons, measured impulse responses can thus be identified with effective couplings,  $[\eff{G}]_{kj}=\gest_k(J_{kj})$. 
As in our simulations synaptic weights are either $J_E$ or $J_I$, the responses $\gest_k(J_E)$ and $\gest_k(J_I)$ have to be measured.
If the synaptic connectivity $J_{kj}$ is known, the coupling matrix is then fully determined. If only certain statistics of the covariances are of interest, knowledge of statistical properties  of the connectivity matrix may be sufficient, see Section \ref{ringSection}. 
In principle, $\gest_k(J)$ is different for each neuron $k$, as variances in input affect the dynamical regime it is in, and has to be determined individually. In random networks, the variance in input is relatively small, and to reduce simulation costs, we assume identical impulse responses for all neurons. The response can then be measured as an average across a small population of probed neurons. 

Because the effective autocovariances $\eff{\Ra}$ are not directly observable, the autocovariance functions of the nodes, $\CM_{kk}$ have to be measured instead. According to \eqref{fullCorrFuncs}, they are linked to the $\eff{\Ra}$ by $C_{kk}(\omega)=\sum_j|[(1-\eff{G}(\omega))^{-1}_{kj}]|^2\eff{y}_{jj}(\omega)$.
Inverting this equation removes the network feedback. Together with the coupling matrix, this allows for a linear prediction of the remaining crosscovariances. 

\begin{figure}[b]
\includegraphics[width=\columnwidth]{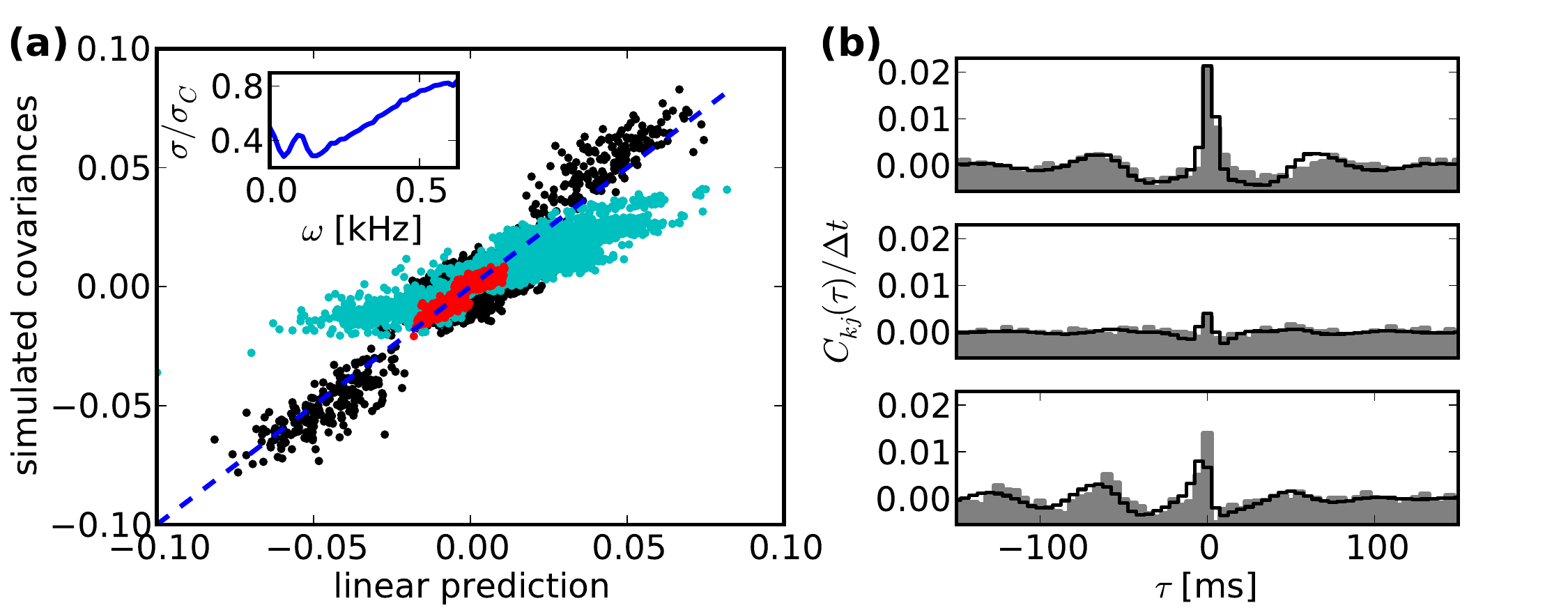}
\caption{(color online) (a) Scatter plot of covariances between neuron pairs. Red (dark gray): $C_{kj}(\omega=0\,\mathrm{kHz})$, black: $\mathrm{Re}[C_{kj}](\omega=0.31\,\mathrm{kHz})$, cyan (light gray): $\mathrm{Im}[C_{kj}](\omega=0.31\, \mathrm{kHz})$. Inset: Standard deviation between prediction and simulation, normalized by the standard deviation of the covariance distribution, increases with frequency.  (b) Typical examples for simulated (gray) and predicted (black) pairwise covariance functions in the time domain. Parameters as in Fig.\ \ref{netwState}.}
\label{linAppOfOm}
\end{figure}

\section{Covariances in random networks}
The linear model then predicts covariance functions between individual neurons in frequency or, equivalently, in time domain, see Fig.\ \ref{linAppOfOm}. Predictions are particularly accurate for lower frequencies.

\begin{figure}
\includegraphics[width=\columnwidth]{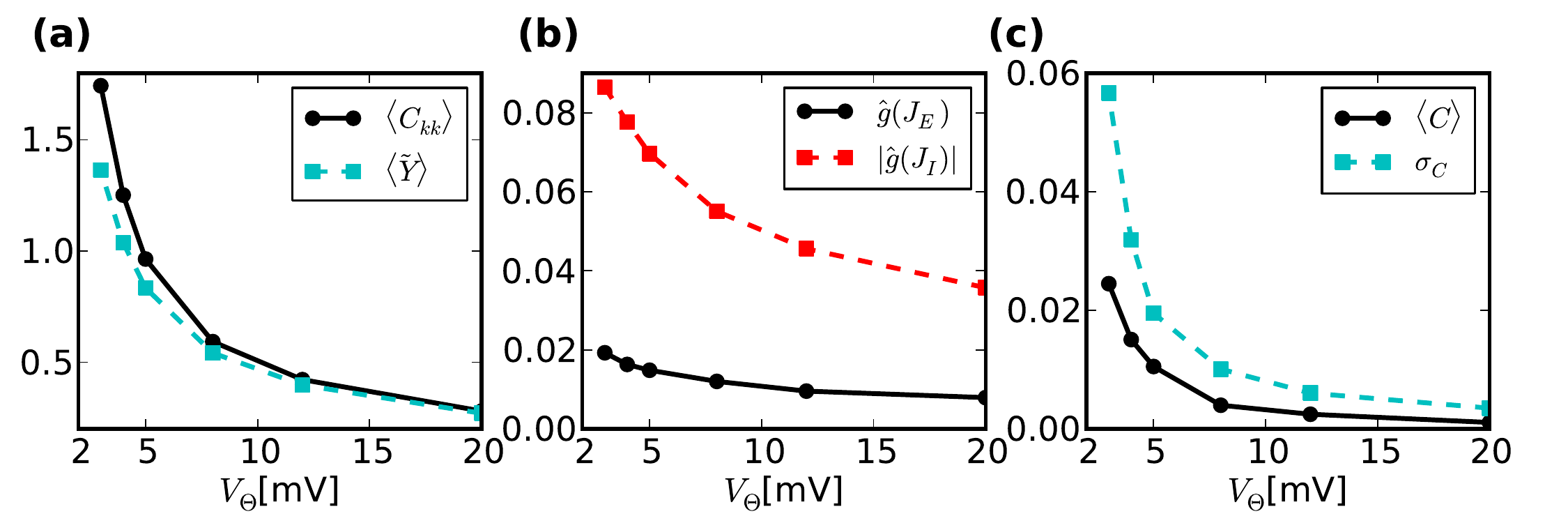}
\caption{(color online) 
Larger reset decreases (a) effective and network autocovariances,
(b) integrated impulse response and (c) mean and standard deviation of distribution of integrated covariances in a random network. 
Parameters as in Fig. \ref{netwState}.
}
\label{linapprox}
\end{figure}

A reset after each spike is a feature of the biophysical mechanism of spike generation, and an integral part of many neuron models. By varying the parameter  $\rv$, the effective resets $\theta_k$ are changed and its consequences for the covariances in the network can be studied.  
In particular, a divergence between $\exval{\eff{\ra}}$ and $\exval{\ra}$ is seen for low values of $V_{\theta}$, that is small resets. This illustrates the point that for the estimation of crosscovariances, it is essential to remove the feedback from the autocovariances, and use the effective values instead.
Simulation results are in line with expectations for the dependence on self-inhibition from the linear theory, Fig.\ \ref{linapprox}.
 An increase of $\rv$  causes weaker impulse responses and smaller autocovariances as well as weaker covariances.

\section{Linear stability} 
A necessary condition for the stability of the linear system \eqref{Hawkesbasic} is that the real part of all eigenvalues of the matrix $\int \KM(\tau)\, d\tau$ is smaller than 1. This condition is also sufficient if the coupling kernels have the shape of a decaying exponential
\cite{Pernice2011}.
We examined how the spectrum of the integrated coupling matrix scales with the size $N$ of a random network, assuming a uniform connection probability $p$.

The  entries of the integrated coupling matrix can be computed analytically \cite{Brunel2000,Helias2010c}. Starting point is the self-consistent rate of the recurrent network $y_{sc}$, which is given by the solution of the self-consistency equation
\be
\frac{1}{y}=\tau_r+\int_{(V_T-\rv-\mu)/\sigma}^{(V_{T}-\mu)/\sigma}e^{u^2}(1+\mathrm{erf}(u))\, du,
\ee
 where $\mu=\tau_m(J_{E}y_{ext}+pN\epsilon J_Ey+pN(1-\epsilon)J_Iy)$ and $\sigma=\tau_m(y_{ext}J_{E}^2+pN\epsilon yJ_E^2+pN(1-\epsilon)yJ_I^2)$ are the mean and the variance of the total input to each neuron. 
For small synaptic weights $J_{E},J_I$, the integrated impulse response to a delta-shaped current is $\gest_{E,I}\equiv\int \gest(J_{E,I},t)dt =\tau_m J_{E,I}\frac{dy}{d\mu}|_{\mu(y_{sc})}$, with
\begin{align}
\frac{dy}{d\mu} & =-y^2\frac{\tau_m}{\sigma}\sqrt{\pi}\Bigl(e^{((V_{T}-\mu)/\sigma)^2}\bigl[1+\mathrm{erf}((V_{T}-\mu)/\sigma)\bigr]\\ \nonumber
 & -e^{(V_T-\rv-\mu)/\sigma^2}\bigl[1+\mathrm{erf}((V_T-\rv-\mu)/\sigma)\bigr]\Bigr).
\end{align}

The spectrum of the adjacency matrix of random networks comprising excitatory and inhibitory neurons has been described previously \cite{Rajan2006}. In the limit of large networks, one eigenvalue corresponds to the average input to each neuron, $m\approx pN(\epsilon\gest_E+(1-\epsilon)\gest_I)$. The remaining eigenvalues are located within a circle in the complex plane of radius $\rho=\sqrt{N(\epsilon \hat{g}_E^2 + (1-\epsilon) \hat{g}_I^2)p(1-p)}$. If the network is balanced or dominated by inhibition, $m\leq 0$ and stability depends only on $\rho$.

\begin{figure}[b]
    \includegraphics[width=\columnwidth]{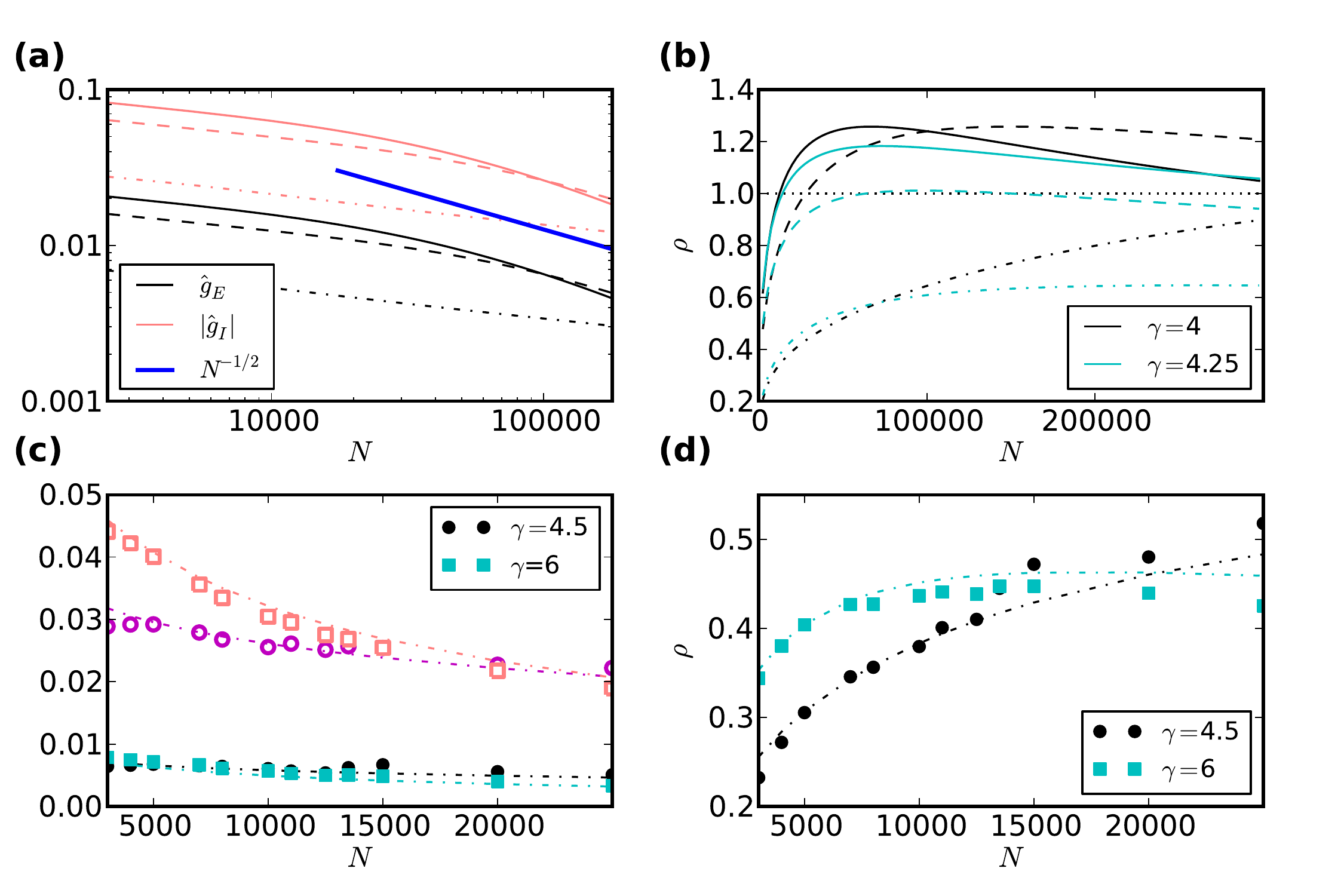}\\
\caption{ (color online) (a) Semi-analytical $\gest_{E,I}$ in random networks of size $N$ for different values of $J_E=0.1,0.4,0.6$ (dot-dashed, dashed, full lines), $\gamma=4$. (b) Bulk spectrum radius $\rho$ of corresponding coupling matrices for $\gamma=4,4.5$. (c) Semi-analytical couplings (dot-dashed lines) and values from simulations with $J_E=0.1$. Circles: $\gamma=4.5$, squares: $\gamma=6$. Empty symbols: excitatory couplings $\gest_{E}$ for $\gamma=4.5, 6$. Filled symbols: corresponding inhibitory couplings $|\gest_{I}|$.  (d) Same for $\rho$.
}
\label{scalingFig}
\end{figure}

Fig.\ \ref{scalingFig} shows that for large system size $\gest_{E,I}$ decreases faster than $N^{-1/2}$ in a balanced network. As $\rho\propto N^{1/2}$, this implies small $\rho$ and guarantees stability and low covariances, in line with previous work \cite{Brunel2000}. In the case of inhibition dominance, the increase of input in larger networks depresses firing rates, causing an even more rapid decay of impulse responses.
However, in intermediate sized networks with weak inhibition and relatively strong synapses, the system can become linearly unstable as $\rho$ grows above 1.
The analytical values for impulse responses and corresponding $\rho$ are in good agreement with simulations for small networks far away from the unstable regime.
Activity in networks in the linearly unstable regime is bounded, as firing rates saturate e.g.\ due to the refractory time, but an asynchronous irregular state turns into a state with strong pairwise covariances, Fig.\ \ref{resStab}(a).  The average integrated covariances reflect this phenomenon, they increase steeply for large $\rho$, panel (b).

\section{Influence of coupling kernels}
\begin{figure}
\includegraphics[width=\columnwidth]{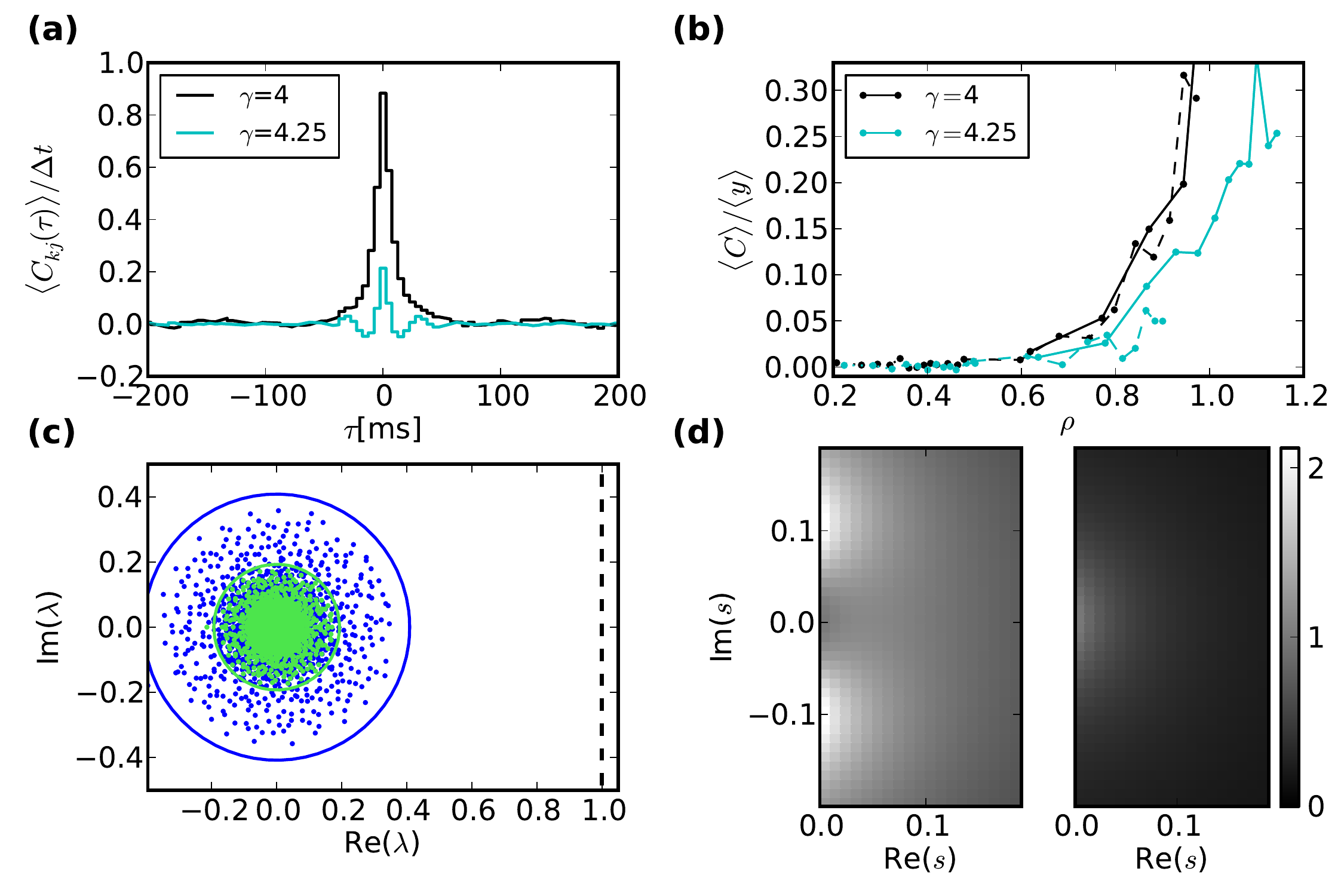}
\caption{
(color online)
(a) Average of covariance functions across pairs ($N=25000$, $J_E=0.4$). 
(b) Integrated covariances normalized by average rates over $\rho(N)$, for $J_E$ as in Fig.\ \ref{scalingFig}. Lines for different $J_E$ partially overlap. Simulation time $10^2\, \mathrm{s}$, other parameters as in Fig.\ \ref{netwState}.
(c) Eigenvalues of integrated coupling matrix (green/light gray) and matrix rescaled by $\gest(s_0)/\gest(0)$ at resonance $s_0$ determined from fit to excitatory kernel in Fig.\ \ref{netwState} (blue/dark gray).
(d) Normalized Laplace transformed impulse response $|\LT{\gest(s)}|/\gest(0)$ for kernel fit $\cos(\nu\tau)e^{-\tau/\alpha}$ (left) and exponential kernel ($\nu=0$).
}
\label{resStab}
\end{figure}

The shape of the coupling kernels is not well described by a decaying exponential, compare  Fig.\ \ref{netwState}.
We therefore take a closer look at the stability of the system described by Eq.\ \eqref{Hawkesbasic}. It can most easily be analyzed, when the spike train $s(t)$ is replaced by the conditional intensity $\ra(t)$. The resulting equation is
\be
\ra(t)=\ra_0+\KM\star\ra(t).
\label{hawkesRate}
\ee
A result from Paley-Wiener theory \cite{gripenberg1990} 
is that a linear system is stable if and only if the real part of all poles of the transfer function, which is defined in the Laplace domain as
\be
\LT{T}(s)\equiv \frac{\LT{y}(s)}{\LT{y_0}(s)}
\ee
and relates input $\ra_0$ and output $\ra$, is smaller than 0. After a Laplace transform of Eq.\ \eqref{hawkesRate} one obtains
\be
\LT{T}(s)=(1-\LT{\KM}(s))^{-1},
\ee
the typical expression for a system with feedback. An alternative formulation for the stability condition is that 1 must not be part of the spectrum of $\LT{G}(s)$ for any $s$ from the right half-plane.
This condition on the spectrum can be readily evaluated, when, for simplicity, we assume that all coupling kernels have the same time dependence, so that excitatory and inhibitory couplings only differ by a scaling factor. The coupling matrix is then of the form $\KM=g(t)K$, where $K$ is a scalar matrix. The bulk spectrum of $K$ is circular and has a radius $\rho_K$. 
The spectrum of  $\LT{G}(s)$ then is a scaled version of the spectrum of $K$, with radius $\rho_K|\LT{g}(s)|$, compare Fig.\ \ref{resStab}(c), as the rotation caused by the complex factor  $\LT{g}(s)$ does not change the shape of the bulk spectrum.
If $\lim_{\operatorname{Re}(s) \to \infty}|\LT{g}(s)|=0$, the bulk spectrum is arbitrarily small, therefore, because 1 must not to be part of it, stability requires that $\rho_K|\LT{g}(s)|<1$ for all $s$ in the right half-plane. If $|g(s)|$ has its maximum at $s=0$, as is the case for exponential kernels, the spectrum of the integrated coupling matrix is decisive, because $\LT \km(0)=\int \km(t)\, dt$.
However, the impulse response of LIF neurons has resonances \cite{Brunel2001a}. We approximate the shape of the impulse response obtained in simulations by the function $\gest(\tau)=ae^{-\tau/\alpha}\cos(\nu \tau)$, compare Fig.\ \ref{netwState}. Resonances appear as maxima in $\LT{\gest}(s)$ for $s\neq 0$, see also Fig.\ \ref{resStab}(d), left panel. These networks are therefore less stable than indicated by the integrated impulse response only. This is not the case for exponential kernels (right panel).

\section{Distance dependent connectivity}
\label{ringSection}
 In \cite{Pernice2011} we demonstrated that,  by expanding Eq.\ \eqref{fullCorrFuncs} into
\be
C(\omega)=\sum_{m,n}\KM^n(\omega)\Ra(\omega)(\KM^*)^m(\omega),
\label{cexpansion}
\ee
covariances can be interpreted as  a sum of contributions from direct and indirect interactions, under the condition that the eigenvalues of $\KM$ have an absolute value smaller than one.
In this series, the terms with $\{n,m\} = \{1,0\}$ describe covariances caused by direct interactions between nodes. The term $\KM(\omega)\Ra(\omega)\KM^*(\omega)$ counts the contribution from common input and terms with higher powers describe indirect interactions via intermediate nodes.
To demonstrate how covariances are connected to these indirect interactions as well as the usefulness of the linear approximation in networks with non-random structure, we consider nodes arranged in a ring, where connection probability between two neurons depends on their geodesic distance $d$. We use a boxcar connectivity function, $p_a(d)=A_a\Theta(d-\sigma_a)$, where $A_a$ and $\sigma_a$ are constants for  $a\in \{E,I\}$, meaning that the connectivity profile is different for excitatory and inhibitory neurons. 

\begin{figure}
\includegraphics[width=\columnwidth]{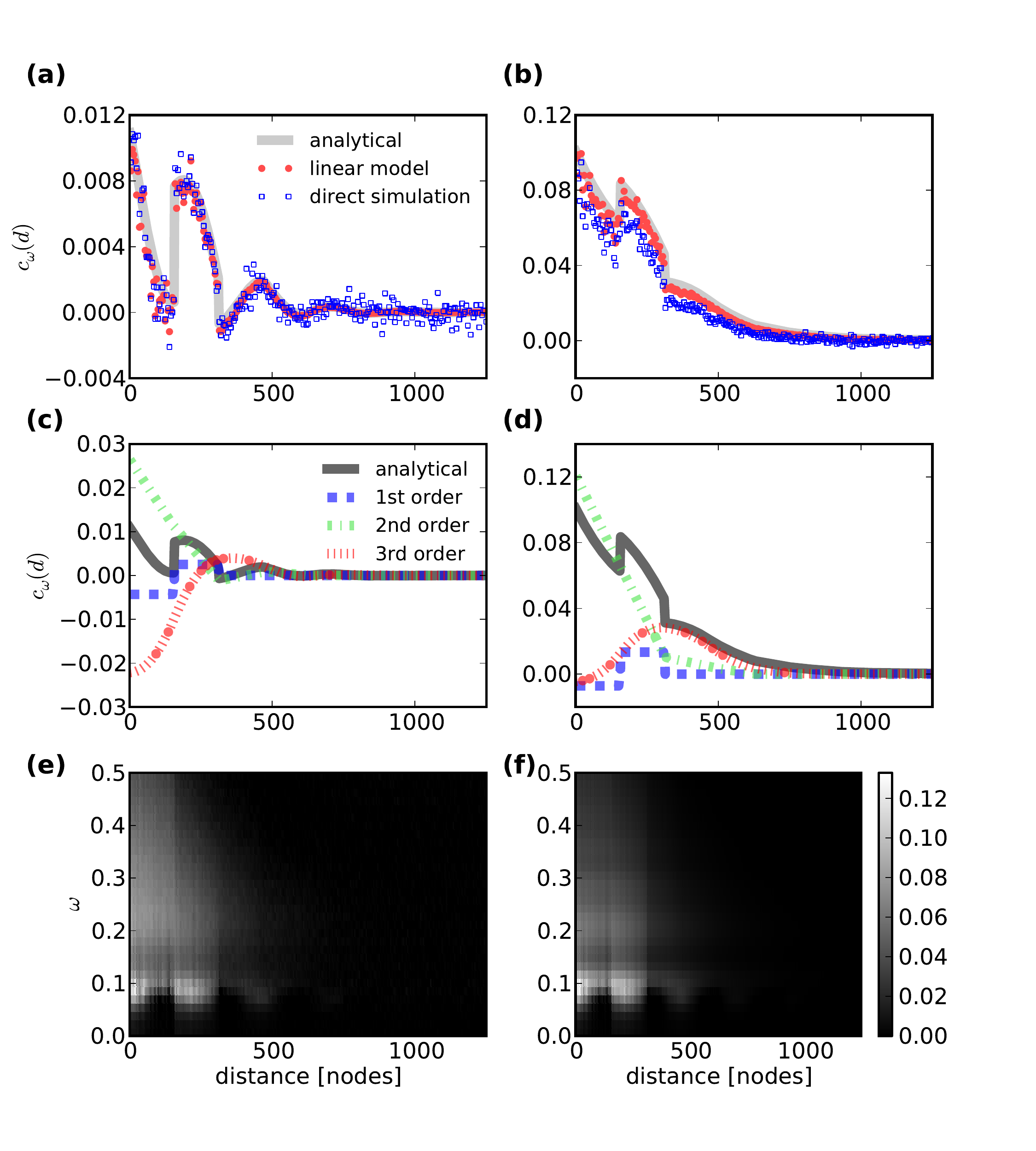}
\caption{
(color online)
Average of covariances over pairs of equal distance for $\omega=0\,\mathrm{kHz}$ (a) and $\omega=0.11\,\mathrm{kHz}$ (real part) (b). Covariances in the direct simulation are well reproduced by the linear approximation. Covariances in the linear model fluctuate around the analytical profile due to the random realization of the network.
Contributions of different orders can be delineated in the analytic expression  \eqref{indContribs} (c),(d).
Distance and frequency dependence of covariance (real part) in direct simulations (e) and the linear approximation (f).
Simulation time was $2\cdot10^4\, \mathrm{s}$, $p=0.05, g=6.5, \sigma_E=312, \sigma_I=156, N=2500$.
}
\label{ringFig}
\end{figure}

In these networks the distance dependent covariance can be defined as $c_{\omega}(d)\equiv\exval{C_{k,k+d}(\omega)}_k$. The quantity $c_{\omega}(d)$ depends on the average coupling $\gamma(d)\equiv\epsilon \gest(J_E,\omega)p_E(d)+(1-\epsilon)\gest(J_I,\omega)p_I(d)$ and the average common input $\eta(d)\equiv(\epsilon\gest(J_E,\omega))^2p_E\star p_E(d)+((1-\epsilon)\gest(J_I,\omega))^2p_I\star  p_I(d)$, where $\star$ denotes a discrete convolution. These expressions result from the positive and negative interactions of the two neuron populations.
Analogously to Eq.\ \eqref{cexpansion}, $c_{\omega}(d)$  then is decomposed in a sum over interactions of different orders, 
\be
c_{\omega}(d)=\sum_{m,n}c_{\omega}^{(m,n)}(d),
\label{indContribs}
\ee
where the matrix products in \eqref{cexpansion} are replaced by convolutions of the average interactions, $c_{\omega}^{(m,n)}=\exval{\eff{y}}\gamma^{\star(n-1)} \star\eta \star \gamma^{\star(m-1)}$ for $m,n>0$ and $c_{\omega}^{(0,1)}=c_{\omega}^{(0,1)}=\exval{\eff{y}}\gamma$.
Application of the spatial Fourier transform
\[
\FT{c_{\omega}}(l)=\sum_{d=0}^{N-1}c_{\omega}(d)\exp{(-i2\pi ld/N)}
\]
leads to the closed expression 
\[
\FT{c_{\omega}}=\exval{\eff{y}}\Bigl( 1+\frac{2\FT{\gamma}}{1-\FT{\gamma}}+\frac{\FT{\eta}}{(1-\FT{\gamma})^2}\Bigr).
\]
For details of the calculation compare \cite{Pernice2011}. 
The effective autocovariance $\exval{\hat{y}}$ does not affect the shape of the distance dependence and is merely a proportionality factor. Application to the simulated LIF networks in Fig.\ \ref{ringFig} shows that the distance dependence of the covariance results from the sum of the direct and many indirect contributions. Both the shape and the spatial extent depend on indirect paths. Higher order terms corresponding to indirect contributions contribute strongly, compared to the direct interactions. 
For covariances  on long time scales, represented by the zero-frequency covariance, positive and negative contributions partly cancel, so that effective covariances are of moderate size. 
 The relative contributions of the higher order terms vary with frequency, due to the frequency dependence of the linear response. As a consequence, the spatial profile of the distance dependence is strongly frequency dependent.

\section{Discussion}
 We have shown that a linear theory describes covariances in networks of LIF neurons in approximately asynchronous states. While linearization is a common approach to approximate complex dynamics, the existence of multiple feedback loops in recurrent networks casts doubts on the applicability of a linear theory. In retrospect, one apparent reason is that contributions of different order effectively cancel due to the interaction of excitatory and inhibitory populations. 

The price to pay to apply linear response theory in this framework is that both the linear response  and the autocovariances of the neurons need to be inferred from  simulations, measurements or an extended analytical framework. 
Even though, information about the dynamics, as the linear stability of the network, the distance dependence of covariances, or the relative contributions of indirect interactions can be inferred from the impulse response and the network structure alone.
An additional asset is the possibility to compare different connectivity structures in a manner that does not depend on the specific neuron model used. So far, this problem has been approached mainly in simulations of the full dynamics of the network \cite{Zhao2011,Yger2011}, but see \cite{Roxin2011}, where a rate model has been used and \cite{Trousdale2011} for an alternative linearized approach.

Single neuron properties determine the impulse response and interact with the network structure to define the overall covariance configuration. In the special case of LIF neurons, impulse responses have been studied in detail \cite{Brunel2001a,Lindner2001}. Generally, the impulse response in a certain dynamical state can be measured from external input in simulations of more complicated neuron models or in experiments. Similar to the way that the fluctuation driven regime seems to favor linear dynamics for LIF neurons, we expect that this holds for networks of other neuron models as well, as long as correlations and population fluctuations are small.

Although our approach is restricted to stationary conditions, it represents a first step towards a better understanding of the influence of the full network structure on the behavior of neuron populations.
One such aspect that is difficult to investigate in mean field models is the stability constraint imposed by the radius of the bulk spectrum of the network related to the topological randomness of the system \cite{Rajan2006}. It implies that global balance of excitation and inhibition is a necessary, but not sufficient condition for stable activity in the network. Simulations in the regime where stability is violated indicate that covariances become large and strong activity fluctuations occur. 
This issue has not been addressed much in the study of neural networks, but it was shown that stability constraints on growing networks can lead to a complex micro-structure \cite{Perotti2009}. Resonances in the impulse responses can equally affect stability, similar as in \cite{Ledoux2011} where bifurcations related to the coupling between populations were studied.

We use networks with constant connection probability which are diluted rather than sparse in the large size limit. For network sizes accessible in simulations, this scaling covers measured values for the estimated number of a few thousand synapses for each neuron in local cortical networks where the connection probability has been estimated to be about 0.1.
We have shown in \cite{Pernice2011} that the  relative impact of more indirect contributions is more significant if the spectral radius of the network is large. In the present work we extend these results and show that, as the  effective coupling strength depends on the frequency, the resulting differences in relative weighting of indirect contributions leads to a complex frequency dependence of spatiotemporal correlations.
In \cite{Roxin2011} it was demonstrated that small changes in the out-degree variance can lead to a synchronous state when the system is close to a bifurcation, indicating that connectivity has a large influence if networks are almost unstable. On a functional level, a network where the largest eigenvalue of the adjacency matrix is equal to one has an optimal dynamic range \cite{Larremore2011}, 
and strong recurrence has been proposed to be a useful condition for cortical computations \cite{stimberg2009}.

\begin{acknowledgments}
We wish to thank Rainer Nagel for 
pointing out reference \cite{gripenberg1990} to us.
 This work was supported by the German Research Foundation (CRC 780, subproject C4) and by the German Federal Ministry of Education and Research (BCCN Freiburg, grant 01GQ0420, and BFNT Freiburg*T\"ubingen, grant 01GQ0830).
\end{acknowledgments}

\bibliographystyle{apsrev4-1}
%

\end{document}